\documentclass[letterpaper,12pt]{article}
\usepackage{amsmath}
\usepackage{amsfonts}
\usepackage{amssymb}
\usepackage{graphicx}
\usepackage{physics}
\usepackage{latexsym}
\usepackage{epsfig}
\usepackage{pstricks}
\usepackage{stmaryrd}
\usepackage{rotating}
\usepackage[english]{babel}
\usepackage{setspace}
\usepackage[utf8]{inputenc}
\usepackage{natbib}
\usepackage[T1]{fontenc}
\usepackage{lmodern}
\usepackage{enumitem}
\usepackage{csquotes}
\usepackage{amsthm}
\usepackage{xcolor}
\usepackage[margin=1in]{geometry}
\usepackage{comment}
\begin{document}
%%%%%%%%%%%%%%%%%%%%%%%%%%%%%%%%%%%%%%%%%%%%%%%%%%%%%%%%%%%%%%
%%%%%%%%%%%%%%%%%%%%%%%%%%%%%%%%%%%%%%%%%%%%%%%%%%%%%%%%%%%%%% 	
\begin{center}	
\begin{Large}
\textbf{Fully Self-Consistent Semiclassical Gravity}\\
\end{Large}
\end{center}

\begin{center}
\begin{large}
R. Muciño, E. Okon, D. Sudarsky and M. Wiedemann\\
\end{large}
\textit{Universidad Nacional Aut\'onoma de M\'exico, Mexico City, Mexico.}\\[1cm]
\end{center}

A theory of quantum gravity consists of a gravitational framework which, unlike general relativity, takes into account the quantum character of matter. In spite of impressive advances, no fully satisfactory, self-consistent and empirically viable theory with those characteristics has ever been constructed. A successful \emph{semiclassical} gravity model, in which the classical Einstein tensor couples to the expectation value of the energy-momentum tensor of quantum matter fields, would, at the very least, constitute a useful stepping stone towards quantum gravity. However, not only no empirically viable semiclassical theory has ever been proposed, but the self-consistency of semiclassical gravity itself has been called into question repeatedly over the years. Here, we put forward a fully self-consistent, empirically viable semiclassical gravity framework, in which the expectation value of the energy-momentum tensor of a quantum field, evolving via a relativistic objective collapse dynamics, couples to a fully classical Einstein tensor. We present the general framework, a concrete example, and briefly explore possible empirical consequences of our model.

%\tableofcontents

\onehalfspacing
%%%%%%%%%%%%%%%%%%%%%%%%%%%%%%%%%%%%%%%%%%%%%%%%%%%%%%%%%%%%%% 
%%%%%%%%%%%%%%%%%%%%%%%%%%%%%%%%%%%%%%%%%%%%%%%%%%%%%%%%%%%%%%
\section{Introduction}
%%%%%%%%%%%%%%%%%%%%%%%%%%%%%%%%%%%%%%%%%%%%%%%%%%%%%%%%%%%%%% 
%%%%%%%%%%%%%%%%%%%%%%%%%%%%%%%%%%%%%%%%%%%%%%%%%%%%%%%%%%%%%%

A theory of quantum gravity aims at providing a gravitational framework that, contrary to general relativity, grants matter an inherently quantum nature. Despite decades of research with commendable progress, no entirely satisfactory, fully self-consistent and empirically viable theory of quantum gravity has been successfully formulated. Various approaches, such as string theory and loop quantum gravity, have been proposed, but none have yet been confirmed through experimental evidence or resolved all of the deep conceptual challenges involved.

A promising intermediary step towards a full theory of quantum gravity is semiclassical gravity, which seeks to bridge the gap between classical and quantum descriptions. In such a framework, the classical Einstein tensor---fully characterizing the structure of spacetime---is coupled to the expectation value of the energy-momentum tensor of quantum matter fields. This approach suggests that, while spacetime remains classical, the gravitational field interacts with matter in a way that reflects its quantum nature. If successfully formulated, semiclassical gravity could serve, at the very least, as a crucial stepping stone toward a deeper understanding of quantum gravity. However, no empirically viable semiclassical theory has been conclusively established, and the fundamental self-consistency of semiclassical gravity has been repeatedly questioned over the years. Some of the main challenges include issues related to self-consistency, empirical adequacy and potential violations of causality.

In this work, we propose a novel semiclassical gravity framework that is both fully self-consistent and empirically viable. Our approach is based on a coupling between the expectation value of a quantum field, which evolves according to a relativistic objective collapse dynamics, and a purely classical Einstein tensor. Objective collapse theories modify standard quantum mechanics by introducing a mechanism that causes wave function collapse in a way that does not require an external observer, thus offering a potential resolution to the quantum measurement problem. By integrating objective collapse into a semiclassical gravitational framework, we construct a theory in which quantum matter influences classical spacetime in a well-defined and consistent manner. Our general attitude is to interpret our model as a potential effective or approximate description, eventually to be surpassed by a full fledged quantum gravity theory. We must say, though, that we do not find any in principle reason for our semiclassical model not to serve as a candidate for a fundamental theory itself.

Below, we present the general theoretical framework underlying our approach, provide a concrete example illustrating its implementation, and briefly discuss potential empirical consequences that could be tested in future experiments. Our results suggest that objective collapse dynamics may offer a viable path toward reconciling quantum mechanics and general relativity, while simultaneously providing new insights into the nature of gravitational interactions at the quantum level.

Our manuscript is organized as follows. In section \ref{SCG}, we describe the semiclassical gravity program and review the main arguments against it. Next, in section \ref{RCM}, we introduce a relativistic, objective collapse framework and explore its potential in addressing shortcomings within semiclassical gravity. In section \ref{SCU}, we present a fully self-consistent semiclassical framework, in which the expectation value of a quantum field, evolving via a relativistic objective collapse dynamics, couples to a fully classical Einstein tensor. Then, in section \ref{Ex}, we develop a concrete application of our general framework and, in section \ref{Tab}, we explore some empirical consequences of our model in recently proposed tabletop experiments. Finally, section \ref{Con} contains our conclusions.

%%%%%%%%%%%%%%%%%%%%%%%%%%%%%%%%%%%%%%%%%%%%%%%%%%%%%%%%%%%%%% 
%%%%%%%%%%%%%%%%%%%%%%%%%%%%%%%%%%%%%%%%%%%%%%%%%%%%%%%%%%%%%%
\section{Semiclassical gravity and its potential problems}
\label{SCG}
%%%%%%%%%%%%%%%%%%%%%%%%%%%%%%%%%%%%%%%%%%%%%%%%%%%%%%%%%%%%%% 
%%%%%%%%%%%%%%%%%%%%%%%%%%%%%%%%%%%%%%%%%%%%%%%%%%%%%%%%%%%%%%

Semiclassical gravity is a formalism in which the structure of spacetime is modeled in general relativistic terms, but matter is treated quantum mechanically. The basic equations of the formalism are given by
\begin{equation}\label{SemiClass}
	G_{ab}=8\pi G \bra{\psi}\hat{T}_{ab}\ket{\psi},
\end{equation} 
where the (renormalized) expectation value of the energy-momentum tensor of the quantum fields acts as the source of the classical Einstein tensor. 

The viability of semiclassical gravity has been called into question time and again over the years, with a variety of influential arguments alleging for inconsistency, the possibility of superluminal signaling or empirical inadequacy. An early, influential argument is developed in \cite{eppley1977necessity}, which introduces a thought experiment in which the position of a quantum particle, assumed to have small uncertainty in momentum, is measured with a classical gravitational wave with small wavelength \textit{and} small momentum. It is then argued that there are three different possible outcomes of the experiment: i) that the measurement localizes the quantum particle to within the gravitational wave, without transferring a large momentum---in which event the uncertainty principle would be violated; ii) that the measurement localizes the quantum particle, without violating the uncertainty principle---in which event energy conservation would be violated; iii) that the measurement does not collapse the wavefunction of the particle---in which event the measurement could be used for superluminal communication.

It has been pointed out, though, that the argument in \cite{eppley1977necessity} has a number of weaknesses. For instance, \cite{huggett2001quantize} explains why the argument fails to deliver a true no-go theorem. Even worse, \cite{mattingly2005quantum} shows that the proposed experiment cannot be carried-out, even in principle. At any rate, tighter objections have been raised since, below we review them in some detail.

%%%%%%%%%%%%%%%%%%%%%%%%%%%%%%%%
\subsection{Inconsistency and empirical inadequacy}
%%%%%%%%%%%%%%%%%%%%%%%%%%%%%%%%

An often cited assessment of the viability and empirical adequacy of semiclassical gravity is carried out in \cite{Page}, in which an actual experiment is reported. The experimental setup employed uses a quantum event (e.g., measurement of the spin along $z$ of a spin-$\frac{1}{2}$ particle prepared spin-up along $x$) to determine whether a sphere of mass $M$ is displaced to the right or left.\footnote{The experimental setup in \cite{Page} is actually more complex, but the details are not relevant for the present discussion.} Then, the gravitational field of the sphere is measured through a Cavendish-like torsion balance. The result, as expected, shows that when the quantum event presents a displacement towards one side, the balance experiences a gravitational field at the corresponding location.

Some advocates of different quantum gravity programs take this enthusiastically as an indication of the non-viability of semiclassical gravity. The idea is that the results allow for two possible conclusions: either semiclassical gravity is internally inconsistent, or it isn't empirically adequate; the reasoning goes as follows.

On the one hand, if there are no quantum collapses, the spin measurement would leave the spin-$\frac{1}{2}$ particle in a spin-up and spin-down superposition, the mass $M$ would get into a superposition of being displaced towards the right and the left and the gravitational effect of the sphere would retain its initial symmetry, so the test mass would not move. This, however, is contradicted by the experimental result. On the other hand, if quantum states do undergo at some times some sort of quantum collapse, then the divergence of the expectation value on the right-hand-side is not generically guaranteed to vanish. However, as a result of the Bianchi identities, the divergence of the left-hand-side of Eq. (\ref{SemiClass}) is always zero. In that case, the framework would be inconsistent. In \cite{Page} this is taken as supporting the notion that the gravitational field must be quantized

As strong as the argument may seem, we do not take it as a conclusive indication that semiclassical gravity is worthless (see \cite{carlip2008quantum} for another defense of the semiclassical program). Indeed, in \cite{Page} to potential problems of the semiclassical framework are brought into the table: either the program is empirically inadequate, especially when large dispersion in $\hat{T}_{ab}$ is present, or the equations are inconsistent, particularly during collapses. However, not everything is lost for the program.

Regarding the first issue, its strength clearly depends upon the quantum framework employed, which determines the extent to which it is possible to construct quantum states with a large enough energy-momentum dispersion to make semiclassical gravity empirically inadequate. For instance, one might try to argue that, due to decoherence, the superposition of $M$ being at two distinct places considered in \cite{Page} never actually realizes. The problem with this line of thought, however, is that the vagueness present in the standard quantum formalism does not really allow for a general, rigorous assessment of the issue. In other words, the ambiguities associated with the collapse postulate in standard quantum mechanics get in the way of deciding if it is actually possible to prepare a macroscopic system in a quantum superposition of empirically relevant distinct positions. In section \ref{RCM} we will explore relativistic collapse theories, which actually permit us to provide a concrete answer to this issue, consistent with the empirical result obtained in \cite{Page}, undermining the conclusion that, in order to account for such experiments, it is necessary to resort to full quantum gravity.

As for the second issue above, we acknowledge that it constitutes a serious hurdle for taking semiclassical gravity as a fundamental description, exactly valid always. That being said, we do not see it as an impediment for looking at it as a viable description, with restricted but rather wide applicability---or even as an exact description, valid only under certain circumstances. In particular, we expect the semiclassical equation to stop operating in some situations or regions, like those associated with quantum collapses.\footnote{In fact, in \cite{maudlin2020status} it is shown that all solutions to the conceptual problems of standard quantum theory entail violations of conservation of the stress-energy tensor.} We take this to be somehow analogous to the standard reading of the Navier-Stokes equations as applied to fluid dynamics. Such equations provide a robust description of fluids under a wide set of circumstances. Yet, they break down under certain conditions, such as when turbulence ensues or during the breaking of a wave in the ocean.

%%%%%%%%%%%%%%%%%%%%%%%%%%%%%%%%
\subsection{Signaling}
%%%%%%%%%%%%%%%%%%%%%%%%%%%%%%%%

Above we mentioned that, according to \cite{eppley1977necessity}, a potential consequence of semiclassical gravity would be the possibility of sending superluminal signals. There is, in fact, a more elaborated and modern argument in this direction, developed in \cite{bahrami2014schrodinger}. The idea is, on the one hand, that the non-relativistic limit of the semiclassical equations is the Schrödinger-Newton equation 
\begin{equation}\label{SNeq}
i \hbar \partial_t \psi(t, \mathbf{r})=\left(-\frac{\hbar^2}{2 m} \nabla^2-G m^2 \int \mathrm{d}^3 \mathbf{r}^{\prime} \frac{\left|\psi\left(t, \mathbf{r}^{\prime}\right)\right|^2}{\left|\mathbf{r}-\mathbf{r}^{\prime}\right|}\right) \psi(t, \mathbf{r}).
\end{equation}
On the other hand, it is argued that the Schrödinger-Newton equation, together with the standard collapse postulate, leads to faster-than-light signaling. The upshot is that semiclassical gravity would seem to allow superluminal communication. 

To argue for the claim that the non-relativistic limit of semiclassical gravity is the Schrödinger-Newton equation, \cite{bahrami2014schrodinger} shows that, in a weak-field limit, with $g_{ab}=\eta_{ab}+h_{ab}$ and in the Newtonian limit, in which $v\ll c$, Eq. (\ref{SemiClass}) reduces to the Poisson equation\footnote{This is analogous to standard general relativity in the Newtonian limit, where the curvature of spacetime is reduced to a Newtonian potential produced by the energy density term of the energy-momentum tensor.}
\begin{equation}\label{Poisson}
	-\frac{c^2}{2}\nabla^2 h_{00}=\frac{4\pi G}{c^2}\langle\psi|\hat{T}_{00}|\psi \rangle .
\end{equation}
In the non-relativistic limit, this potential contributes to the Hamiltonian of the matter fields, yielding the non-linear interaction term in the Schrödinger-Newton equation.

The argument for superluminal signaling within the Schrödinger-Newton scheme runs as follows. We first note that the interaction term on the right hand side of Eq. (\ref{SNeq}) implies a \textit{self-attraction} of wave packets. In particular, the two components of a single particle in a superposition of two different locations would attract each other. With this in mind, consider an EPR-type scenario, where a pair of particles in a singlet state are sent to Alice and Bob. Then, Alice measures the spin of her particle in either the $z$ or $x$ direction and, finally, Bob measures his particle along $z$. 

Now, assuming a standard quantum collapse during measurements, Bob's particle would end up in different states, depending on Alice's choice of measurement and outcome. If Alice chooses $z$, Bob's particle would collapse to either $\ket{z+}$ or $\ket{z-}$, in which case Bob's particle would end up with some displacement $\pm d$ from the center of the detection screen. However, if Alice chooses the $x$ orientation, Bob's particle would enter the apparatus in a superposition of $\ket{z+}$ and $\ket{z-}$. In that case, the two wave packets would attract, and the detection would show a displacement by a smaller amount $\pm d'$. Thus, Bob would be able to know what Alice decided to measure---and, thus, to receive a superluminal message from her.

This argument could seem quite strong, but it is easy to see that it is powerless. The issue is that, once the Newtonian limit is taken (i.e., $c \rightarrow \infty$), then there is no reason to expect a prohibition of superluminal communication. In fact, it is of course the case that the non-relativistic limit of general relativity is Newtonian gravity, so an argument fully analogous to the one in \cite{bahrami2014schrodinger} would advance that general relativity allows for superluminal signaling, since Newtonian gravity does. Needless to say, such an argument would be absurd; and if that argument cannot be taken seriously, neither can the one under consideration.

Still, this does not show that semiclassical gravity could never allow for superluminal signaling. To address this issue, we point out that, in line with \cite{Page}, in order to make semiclassical gravity empirically viable, some sort of deviation of purely unitary evolution is required. However, as soon as that is the case, the expectation value of the energy-momentum tensor at a point $x$, the right hand side of Eq. (\ref{SemiClass}), is no longer independent of the hypersurface $\Sigma$ containing $x$, from which one reads the quantum state $\ket{\psi}$. Therefore, without a well-defined recipe for choosing $\Sigma$, one does not yet possess a well-defined framework. Another way so state this is to note that, in order to have a well-defined theory, one must be clear about what are the elements of the ontology (or local beables, in the language of Bell) postulated to be the source of gravity. And, if the framework is to be Lorentz invariant, the recipe for defining such local beables better share this invariance. 

In this regard, for this purpose, below we adopt the concrete prescription in \cite{bedingham2011relativistic}, which defines the energy-momentum density tensor at $x$ as the expectation value of the energy-momentum operator, on the state over the past null cone of $x$. Such a choice of hypersurface to compute the expectation value is not only Lorentz invariant, but it fully obstructs any possibility of superluminal signaling. To see this, consider again the scenario proposed in \cite{bahrami2014schrodinger}. With the past null cone recipe in place, the measurement of Alice would be unable to superluminally affect the gravitating local beables on Bob's side, so no superluminal signal could be sent.

%%%%%%%%%%%%%%%%%%%%%%%%%%%%%%%%%%%%%%%%%%%%%%%%%%%%%%%%%%%%%% 
%%%%%%%%%%%%%%%%%%%%%%%%%%%%%%%%%%%%%%%%%%%%%%%%%%%%%%%%%%%%%%
\section{Relativistic collapse models}
\label{RCM}
%%%%%%%%%%%%%%%%%%%%%%%%%%%%%%%%%%%%%%%%%%%%%%%%%%%%%%%%%%%%%% 
%%%%%%%%%%%%%%%%%%%%%%%%%%%%%%%%%%%%%%%%%%%%%%%%%%%%%%%%%%%%%%

Non-relativistic objective collapse theories are extremely successful in dealing with the conceptual problems of standard quantum mechanics. In the relativistic front, even though objective collapses might seem fundamentally incompatible with relativity, steady work over the years has removed most (if not all) barriers, to the point that relativistic collapse theories can be argued to offer the most promising avenue towards a fully relativistic quantum framework, addressing the conceptual limitations of standard quantum theories.

Striving to produce a single evolution equation that incorporates, both, unitary evolution and the standard collapse process, non-relativistic collapse models add non-linear, stochastic terms to the standard Schrödinger equation. The aim, in particular, is to construct a formalism that is empirically indistinguishable from standard quantum mechanics at the micro-scale, but in which unwanted macroscopic superpositions, such as Schrödinger-cat states, are effectively suppressed---all this without having to rely on vaguely defined notions, such as measurements or observers.

GRW, the simplest non-relativistic collapse model \cite{GRW}, stipulates all elementary particles to randomly be subjected to spontaneous localization events around positions selected through a probability law similar to the Born rule. In order to avoid an infinite increase in energy, these localizations are not into exact points, but into small regions around them (with the size of such regions determined by a new fundamental parameter). To reproduce standard quantum mechanics at the micro-scale, the frequency of localization for each individual particle is postulated to be extremely small. However, the dynamics is such that the localization frequency is naturally amplified by increasing the number of particles, leading to an extremely large collapse frequency for macroscopic objects. The upshot is the desired suppression of unwanted macroscopic superpositions.

The Continuous Spontaneous Localization (CSL) model \cite{CSL}, substitutes the abrupt GRW jumps with a continuous, random and non-unitary evolution. Specifically, it supplements the Schrödinger equation with non-linear, stochastic components that drive any initial wave function toward one of the eigenstates of the so-called collapse operator, with probabilities approximating those determined by the Born rule. Typically, the collapse operator is linked to position or mass density, as these choices help prevent superpositions of large objects being located in different places, offering a resolution to the measurement problem.

It is important to be explicit about the \emph{physical interpretation} of objective collapse theories. As should be clear from the discussion above, to address the measurement problem, these theories avoid using the standard interpretation of the quantum state, based on the collapse postulate and the Born rule. By discarding this probabilistic interpretation, without offering a replacement, they would fail to provide a complete physical theory that could make testable predictions. This is because there would be no link established between the mathematical framework and the physical reality the theory aims to explain. Consequently, objective collapse models need an alternative interpretation of the quantum state. One option in this regard, proposed in \cite{Bel:87}, is to interpret the GRW jumps as the entities, or \emph{flashes}, out of which physical objects are made out of. Another option is to read the theory as describing a mass density field, defined through the expectation value of the mass density operator.

The initial objective collapse models described above are non-relativistic. However, fully relativistic versions have also been constructed. For instance, \cite{Tumulka} introduces a relativistic GRW for $N$ non-interacting Dirac particles. Likewise, \cite{bedingham2011relativistic} introduces a relativistic collapse dynamics for a quantum field. Next, we describe the general framework for this dynamics.

Following \cite{bedingham2011relativistic}, to present this relativistic collapse dynamics we employ the interaction picture, in which each spacelike hypersurface $\Sigma$ is assigned a quantum state of the matter fields, $\ket{\Psi_\Sigma}$. The change to the state on $\Sigma$, as one moves to the future along an arbitrary foliation of spacetime, is governed by the Tomonaga-Schwinger equation
\begin{equation}
\label{ST}
i \frac{\delta \ket{\Psi_\Sigma}}{\delta \Sigma(x)} = \mathcal{H}_{\text{int}} (x) \ket{\Psi_\Sigma} ,
\end{equation}
with $\mathcal{H}_{\text{int}} (x)$ the interaction Hamiltonian density. Covariance, in the form of foliation-independence for the dynamics, is guaranteed by the requirement that $[\mathcal{H}_{\text{int}} (x), \mathcal{H}_{\text{int}} (y)]=0$ whenever $x$ and $y$ are spacelike separated.

Mimicking GRW, on top of this unitary dynamics, we stipulate the state to experience discrete collapses associated
to randomly selected spacetime points. In particular, when an hypersurface $\Sigma$ passes through a collapse point $x$, for a moment, the state stops satisfying the Tomonaga-Schwinger equation and evolves according to
\begin{equation}
\label{CD}
\ket{\Psi_\Sigma} \rightarrow \ket{\Psi_{\Sigma^+}} = \hat{L}_x (Z_x) \ket{\Psi_\Sigma}
\end{equation}
with $\hat{L}_x$ the collapse operator at point $x$ and $Z_x$ a random variable selected according to the probability distribution 
\begin{equation}
\label{ProbZ}
P(Z_x|\ket{\Psi_\Sigma}) = \frac{\bra{\Psi_\Sigma} |\hat{L}_x (Z_x) |^2 \ket{\Psi_\Sigma} }{\braket{\Psi_\Sigma}},
\end{equation}
with the collapse operators satisfying 
\begin{equation}
\label{Comp}
\int dZ |\hat{L} (Z) |^2 =1 .
\end{equation}
It is typically assumed that there is a constant probability of collapses per unit 4-volume, a distribution defined in a covariant manner and not depending on any specific choice of foliation (the possibility for the collapse rate to depend on the local curvature of spacetime has also been considered; see \cite{okon2014benefits}).

Now, if the collapse operators satisfy
\begin{equation}
\label{MC}
[\hat{L}_x (Z_x), \hat{L}_y (Z_y)]=0, \quad \text{and} \quad [\hat{L}_x (Z_x), \mathcal{H}_{\text{int}} (y)]=0
\end{equation}
for spacelike separated $x$ and $y$, given hypersurfaces $\Sigma_i$ and $\Sigma_f$, with no point in $\Sigma_i$ to the future of $\Sigma_f$, then the dynamics assigns a foliation-independent state to $\Sigma_f$, with the probability of the complete set of $Z$s between $\Sigma_i$ and $\Sigma_i$, given the state on $\Sigma_i$, independent of the foliation. Then, as long as one finds collapse operators satisfying the above requirements, one has a covariant, foliation-independent framework for a relativistic collapse model.\footnote{See \cite{Myrvold} for a thorough discussion of probabilities in relativistic contexts.} 

A first proposal for the collapse operators is given by
\begin{equation}
\hat{L}_x (Z_x) = \frac{1}{(2 \pi \sigma^2)^{1/4}} \exp{- \frac{ \left( \hat{A}(x)-Z_x \right)^2}{4\sigma^2}},
\end{equation}
with $\hat{A}(x)$ an hermitian operator and $\sigma$ a new fundamental constant. These operators automatically satisfy Eq. (\ref{Comp}) and, since they are quasi projections into approximate eigenstates of $\hat{A}(x)$, the cumulative effect of many collapses would be to lead the system into eigenstates of $\hat{A}(x)$.

This choice of collapse operator satisfies Eqs. (\ref{MC}). However, if one attempts to compute the expected change in energy as the hypersurface $\Sigma$ passes through a collapse point, one encounters an infinite result. This divergence arises because the point-like nature of the collapse introduces spatial discontinuities in the field configuration. For this reason, it seems necessary to introduce a smearing function. There are, however, two potential problems with doing so. The first is that it is not clear how to introduce a smearing function in a Lorentz covariant way. This problem can be overcome by constructing the smearing function at a point, employing either local properties of the state of the field or the local spacetime geometry (see \cite{bedingham2011relativistic,BedSud} for details). The second problem is that, once an appropriate smearing function is introduced, Eqs. (\ref{MC}) might no longer be satisfied, potentially damaging the Lorentz invariance of the framework. An ingenious way to address this is by the introduction of a non-standard, auxiliary field (see \cite{bedingham2011relativistic} for details).

As for the nature of $\hat{A}(x)$, at the non-relativistic level, we saw that the mass density operator plays a central role. In the relativistic context, this suggests for the collapse operator to be associated with the energy-momentum tensor. If so, the dynamics would seem able to evade large uncertainties in the expectation value of $\hat{T}_{ab}$ which, as discussed above, could lead to conflict with observations. Since different components of the energy-momentum tensor operator do not commute among themselves, one can take $\hat{A}(x)$ to be given by some scalar constructed out of $\hat{T}_{ab}$. Alternatively, one could consider energy-momentum itself as the collapse operator, meaning that, once given a coordinate chart, the various components of the energy-momentum tensor would play the role of a set of collapse operators. This, of course, would have to be done in a covariant manner and taking into account the fact that these various components do not commute. That is, such recipe, although apparently very natural, forces one to consider extending the class of collapse dynamics beyond the schemes that have been considered so far, involving, at most, a set of mutually commuting operators.\footnote{A scheme of this sort can be expected to display a behavior analogous to simple situations, where a single collapse operator does \emph{not} commute with the Hamiltonian. In that case, the dynamics, rather than leading the state towards an eigenstate of the collapse operator, takes the system into a quasi-stable condition, where the state moves back and forth in the vicinity of one of the eigenstates of the collapse operator.}

Finally, as with the non-relativistic models, in order to complete the description of this framework, it is necessary to be explicit about its ontology. In this regard, following \citep{bedingham2011relativistic,bedingham2014matter}, we read the theory as postulating the existence in spacetime of an energy-momentum density, $\mathcal{T}_{ab}(x)$, defined through the renormalized expectation value of the energy-momentum tensor operator, calculated in the state over the past null cone of $x$. That is
 \begin{equation}\label{matterdensityeq}
	\mathcal{T}_{ab}(x) \equiv \langle \psi | \hat{T}_{ab} | \psi \rangle^{\text{Ren}}_{\partial J^-(x)} , 
\end{equation}
with $\partial J^-(x)$ the past null cone of $x$; importantly, such a prescription is explicitly Lorentz-invariant. It might be objected that, since the past null cone is not a Cauchy surface, in general, the state is not defined there, so the recipe is not well-defined. Note, however, that instead of considering the state on $\partial J^-(x)$, one could consider any hypersurface going through $x$ and calculate the expectation value with a non-physical state on it, constructed by only considering collapses that happen within the causal past of $x$. 

It is very important to note that, given that the dynamics under consideration includes objective collapses, it is not always the case that $\nabla^a\mathcal{T}_{ab}(x)=0$. In particular, such an equation is valid everywhere, except on the collapse events and their future null cones. In the next section, we explain how all this can be employed to construct a self-consistent semiclassical framework.

%%%%%%%%%%%%%%%%%%%%%%%%%%%%%%%%%%%%%%%%%%%%%%%%%%%%%%%%%%%%%% 
%%%%%%%%%%%%%%%%%%%%%%%%%%%%%%%%%%%%%%%%%%%%%%%%%%%%%%%%%%%%%%
\section{Fully self-consistent semiclassical gravity}
\label{SCU}
%%%%%%%%%%%%%%%%%%%%%%%%%%%%%%%%%%%%%%%%%%%%%%%%%%%%%%%%%%%%%% 
%%%%%%%%%%%%%%%%%%%%%%%%%%%%%%%%%%%%%%%%%%%%%%%%%%%%%%%%%%%%%%

In this section we put forward a semiclassical framework, in which the expectation value of the energy-momentum tensor of a quantum field, evolving according to the relativistic objective collapse dynamics described above, acts as source of the classical Einstein tensor. The expectation value at point $x$ is to be taken in the state of the field over $\partial J^-(x)$. Given that the evolution of the field is punctuated by deviations of unitarity, the expectation value of the energy-momentum tensor is not conserved, neither where collapses occur nor, given the way in which the expectation value is calculated, in their causal future. We then take the semiclassical Einstein equations to hold only in patches, with a well-defined recipe for gluing them, in order to form what we call a Semiclassical Collapse Universe (SCU). Before defining such a construction in detail, we explore a previous work that serves as inspiration. 

%%%%%%%%%%%%%%%%%%%%%%%%%%%%%%%%
\subsection{Motivation: Semiclassical Self-consistent Configurations}
%%%%%%%%%%%%%%%%%%%%%%%%%%%%%%%%

A key motivation for our proposal is the notion of a \textit{Semiclassical Self-consistent Configuration} (SSC), introduced in \cite{P6}. The definition of a SSC is the following: the set $\lbrace g_{ab}(x),\hat{\varphi}(x), \hat{\pi}(x), {\cal H}, \vert \xi \rangle \in {\cal H} \rbrace$ is a SSC if and only if $\hat{\varphi}(x)$, $\hat{\pi}(x)$ and $ {\cal H}$ constitute a quantum field theory for the field $\varphi(x)$ over a spacetime with metric $g_{ab}(x)$, and the state $\vert\xi\rangle \in {\cal H}$ is such that
\begin{equation}\label{scc}
G_{ab}[g(x)]=8\pi G\langle\xi\vert \hat{T}_{ab}[g(x),\hat{\varphi}(x)]\vert\xi
\rangle ,
\end{equation}
with $G_{ab}[g(x)]$ the Einstein tensor of metric $ g_{ab}$ and $\langle\xi| \hat{T}_{\mu\nu}[g(x),\hat{\varphi}(x)]|\xi\rangle$ the expectation value, in the state $\vert\xi\rangle$, of the renormalized energy-momentum tensor of $\hat{\varphi}(x)$ over a spacetime with metric $ g_{ab}$.

On top of this construction, \cite{P6} attempts to introduce objective collapses of the quantum matter field. Such collapses would produce a discontinuous change in the expectation value of the energy-momentum tensor which, in turn, would lead to a modification of the spacetime metric. However, this would produce a change in the Hilbert space to which the state belongs, so a collapse would induce a transition from one SSC to another. Therefore, the proposal is for two different SSCs to describe the situation before and after a collapse, with a suitable collapse hypersurface joining them. That is, one starts with an initial SSC1 and adds rules to select a spacelike collapse hypersurface $\Sigma_C$ and a post-collapse state (in a new Hilbert space), with which one constructs SSC2. SSC2 is required to posses a hypersurface isometric to $\Sigma_C$, which serves as the hypersurface in which SSC1 and SSC2 are joined to form a global spacetime. 

As interesting and suggestive as it is, it is clear that, in order to get off the ground, the above proposal demands for a long list of largely arbitrary selections to be made. For instance, how is $\Sigma_C$ to be chosen? How does a collapse selects a state in a new Hilbert space? How are two SSCs to be glued? Moreover, while collapses are assumed to happen at spacetime points, this proposal somehow associates with them a whole hypersurface $\Sigma_C$. 

In \cite{juarez2023initial}, a concrete recipe for gluing two SSCs is introduced. The idea is to look for the ``smallest jump possible'' between them, by requiring continuity on $\Sigma_C$ for, both, the induced $3$-metric, $g_{ij}$ and the transverse traceless term of the extrinsic curvature. This, together with the constraints, uniquely determines the remaining part of the post-collapse extrinsic curvature. As we explain below, our new proposal offers significant theoretical improvements over this proposal.

%%%%%%%%%%%%%%%%%%%%%%%%%%%%%%%%
\subsection{Semiclassical Collapse Universes (SCUs)}
%%%%%%%%%%%%%%%%%%%%%%%%%%%%%%%%

The central construction of our framework is what we call a Semiclassical Collapse Universe or SCU. In short, a SCU consists of a quantum field theory, with an objective collapse dynamics, living on spacetime $(M, g_{ab})$, with the metric $g_{ab}$ satisfying 
\begin{equation}
\label{SCE}
G_{ab}( x) = 8\pi G \mathcal{T}_{ab}(x)
\end{equation}
everywhere, except on the collapse points and their future light cones, where $g_{ab}$ is discontinuous, but the induced 3-metric is continuous.

In more detail, a SCU consists of a set $\lbrace M, g_{ab}(x), \lbrace x_i,Z_i \rbrace, \hat{L}_i (Z_i ), \mathcal{S} \rbrace$ with $M$ a 4-d manifold, $g_{ab}(x)$ a Lorentzian metric smooth by patches, $\lbrace x_i,Z_i \rbrace$ a set of points $x_i \in M$, each with an associated random variable $Z_i$, $\hat{L}_i (Z_i )$ a set of collapse operators satisfying Eqs. (\ref{Comp}) and (\ref{MC}) and $\mathcal{S}: \mathcal{F} \to \mathcal{H} $ an assignment of a quantum state of the matter fields, $\ket{\Psi_\Sigma} \in \mathcal{H}$, to each spacelike hypersurface $\Sigma \in \mathcal{F} $ (with $\mathcal{F} $ the set of all hypersurfaces in $M$). The set of points $\lbrace x_i \rbrace$ is distributed over $M$ with a constant mean number of elements per unit 4-volume of the metric $g_{ab}(x)$, the random variables $Z_i$ are consistent with the probability distribution in Eq. (\ref{ProbZ}) and $\mathcal{S}$ consistent with the collapse dynamics given by the Tomonaga-Schwinger type equation (\ref{ST}), punctuated with collapse events, where Eq. (\ref{CD}) is valid. Finally, Eq. (\ref{SCE}) is valid everywhere, except on the collapse points $x_i$ and their future light cones, where the induced 3-metric is continuous.

In the next subsection we give more details about this last gluing condition and, in particular, explain why it is sufficient to fully determine the metric everywhere.

%%%%%%%%%%%%%%%%%%%%%%%%%%%%%%%%
\subsection{Gluing conditions}
%%%%%%%%%%%%%%%%%%%%%%%%%%%%%%%% 

To explain the demand for the induced 3-metric to be continuous, we focus on the simple case where only one collapse occurs at $x^\mu$, dividing the SCU into two patches, $J^+(x^\mu)$ and $\mathcal{M}-J^+(x^\mu)$. In a similar spirit to \cite{juarez2023initial}, we impose gluing conditions on $\partial J^+(x^\mu)$ that ensure the ``smallest jump possible'' for the metric and its derivatives, consistent with the constraints and the collapse theory.

To this end, we explore the initial value formulation on null hypersurfaces, which presents important differences compared to the more familiar spacelike case. The most complete treatment for the situation where the initial hypersurface is a future null cone, $C_O$, with vertex at $O$, is given in \cite{choquet2011cauchy}; many of our considerations regarding the gluing conditions are inspired by that work. % (see Appendix \ref{SCB} for a summary).
The key result is that if one prescribes on $C_O$ as initial data a quadratic form, $\bar{g}$, which satisfies a set of four equations known as wave-map gauge constraints, $\mathcal{C}_\alpha=T_{\alpha\beta}l^\beta|_{C_O}$, then the Einstein equation admits a solution $g$ in a neighborhood of the vertex, such that $g|_{C_O}=\bar{g}$. Uniqueness is demonstrated only for the vacuum case, but it is reasonable to assume that, for sensible energy-momentum tensors, the solution should also be unique.

Furthermore, and of special relevance to our analysis, \cite{choquet2011cauchy} studies a method for constructing initial data that satisfies the wave-map constraints, given a 3-metric, $\tilde{g}$, in $C_O$. To achieve this, a special coordinate system $\{x^{\prime\alpha}\}$ is introduced, where $C_O$ is represented by the equation $x^{\prime 0}=(x^{\prime 1})^2+(x^{\prime 2})^2+(x^{\prime 3})^2$. This coordinate system can be constructed using Riemann normal coordinates at $O$, with an orthogonal frame \cite{choquet2009general}. Based on $\{x^{\prime\alpha}\}$, null adapted coordinates $\{x^\alpha\}$ are defined, in which the initial data take the form
\begin{equation}
\bar{g}:=g|_{C_0}:=\bar{g}_{00}(dx^0)^2+2\nu_0dx^0dx^1+2\nu_Adx^0dx^A+\underbrace{\bar{g}_{AB}dx^Adx^B}_{\tilde{g}}.
\end{equation}

Working in this coordinate system, if an admissible 3-metric $\tilde{g}=\bar{g}_{AB}dx^Adx^B$ on $C_O$ is given, where by ``admissible'' is meant that $\tilde{g}_{\alpha^\prime\beta^\prime}(O)=\eta_{\alpha^\prime\beta^\prime}$, it is shown that the wave-map constraints (for reasonable $T_{ab}$) turn into four hierarchical, ordinary differential equations for the remaining four coefficients of $\bar{g}$, all linear, once the physical constraint $(G_{rr}-T_{rr})|_{C_O}=0$ has been solved. Therefore, if one prescribes $\tilde{g}$ and imposes initial conditions for the differential equations such that $\bar{g}_{\alpha^\prime\beta^\prime}(O)=\eta_{\alpha^\prime\beta^\prime}$, one can uniquely determine the remaining data $(\bar{g}_{00}, \nu_0, \nu_A)$ of $\bar{g}$, which guarantees the existence of a solution.

Based on this, our proposal is to impose for the induced 3-metric to remain continuous on $\partial J^+(z^\mu)$. Then, the four wave-map constraints, together with the corresponding jump in the expectation value of the energy-momentum tensor, uniquely determine the remaining coefficients of $\bar{g}$ that ensure the existence of a solution of the Einstein equations in $J^+(z^\mu)$. Note that, since the induced 3-metric in our proposal comes from a physical metric, it is always admissible. 

Before moving on, we must mention that the fact that the metric of an SCU is discontinuous may bring with it technical issues that need to be looked at closely. In particular, the whole construction of a quantum field theory is (reasonably) under control in smooth spacetimes, but the situation for discontinuous ones is much less clear. We leave this analysis for future work.

%%%%%%%%%%%%%%%%%%%%%%%%%%%%%%%%%%%%%%%%%%%%%%%%%%%%%%%%%%%%%% 
%%%%%%%%%%%%%%%%%%%%%%%%%%%%%%%%%%%%%%%%%%%%%%%%%%%%%%%%%%%%%%
\section{Example}
\label{Ex}
%%%%%%%%%%%%%%%%%%%%%%%%%%%%%%%%%%%%%%%%%%%%%%%%%%%%%%%%%%%%%% 
%%%%%%%%%%%%%%%%%%%%%%%%%%%%%%%%%%%%%%%%%%%%%%%%%%%%%%%%%%%%%%

As a concrete realization of our framework, we explore a spherically symmetric setting, which allows us to explicitly solve the equations. In particular, we consider a scenario with a quantum state that gives rise, via Eq. (\ref{matterdensityeq}), to a central massive core with radius $R_C$ and uniform density $\rho_C$, and a massive shell of inner radius $R_I$, outer radius $R_E$ and uniform density $\rho_S$ (see Figure 1).
%%%%%%%%%%%%%%%%%%%%%%%%%%%%%%
\begin{figure}%[ht]
\centering

\includegraphics[height=11cm]{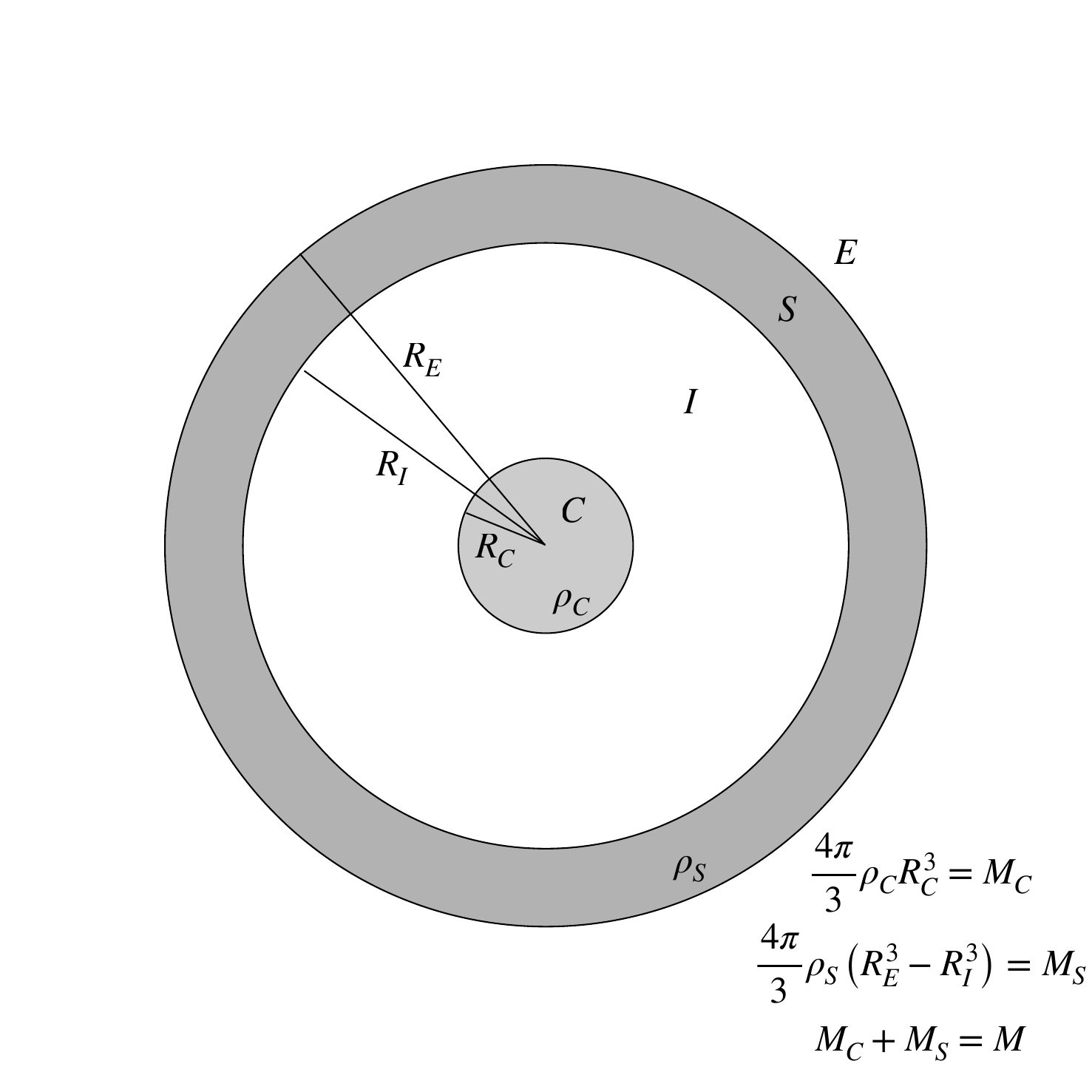} 
\caption{A central massive core with radius $R_C$ and uniform density $\rho_C$, and a massive shell of inner radius $R_I$, outer radius $R_E$ and uniform density $\rho_S$.}
\end{figure}
%%%%%%%%%%%%%%%%%%%%%%%%%%%%%% 
We name the regions the core ($C$), the interior ($I$), the shell ($S$) and the exterior ($E$), and we define
\begin{equation}
\label{Mass}
 M_C = \frac{4\pi}{3} R_C^3 \rho_C , \qquad M_S = \frac{4\pi}{3} \left( R_E^3-R_I^3 \right) \rho_S \qquad \text{and} \qquad M = M_C + M_S.
\end{equation}

For a static, spherically symmetric scenario, the metric everywhere can be written as
\begin{equation}
ds^2 = -e^{2 \phi(r)} dt^2 + \left(1-\frac{2m(r)}{r} \right)^{-1} dr^2 + r^2 d\Omega^2 ,
\end{equation}
with 
\begin{equation}
m(r) = 4 \pi \int_0^r \rho(r') r'^2 dr'
\end{equation}
and
\begin{equation}
\frac{d\phi(r)}{dr} = \frac{m(r) + 4 \pi r^3 P(r)}{r [r-2 m(r)]} ,
\end{equation}
with $P(r)$ the radial pressure (see appendix \ref{EST} for expressions for $m(r)$ and $\phi(r) $ in each region). 

Now, given the relations in Eqs. (\ref{Mass}), we can write the metric everywhere as a function of $\rho_C$, keeping $M$ constant. Moreover, we can transform into null coordinates, with $u=t-\int (1-2m(r)/r)^{1/2} e^{-\phi(r)} dr$, in which case, the metric can generically be written as
\begin{equation}
\label{Mrc}
ds^2(\rho_C) = -e^{2 \phi(r,\rho_C)} du^2 - 2 e^{\phi(r,\rho_C)} \left(1-\frac{2m(r,\rho_C)}{r} \right)^{-1/2} du dr + r^2 d\Omega^2 . 
\end{equation}

Next, we consider an objective collapse which happens to occur exactly at $r=0$ and $t=0$ and localizes all the mass on the shell. Given Eq. (\ref{matterdensityeq}), the effect of such a collapse would travel along the causal future of the collapse event (i.e., $u=0$) and produce the mass distribution depicted in Figure 2. 
%%%%%%%%%%%%%%%%%%%%%%%%%%%%%%
\begin{figure}%[ht]
\centering
\includegraphics[height=10cm]{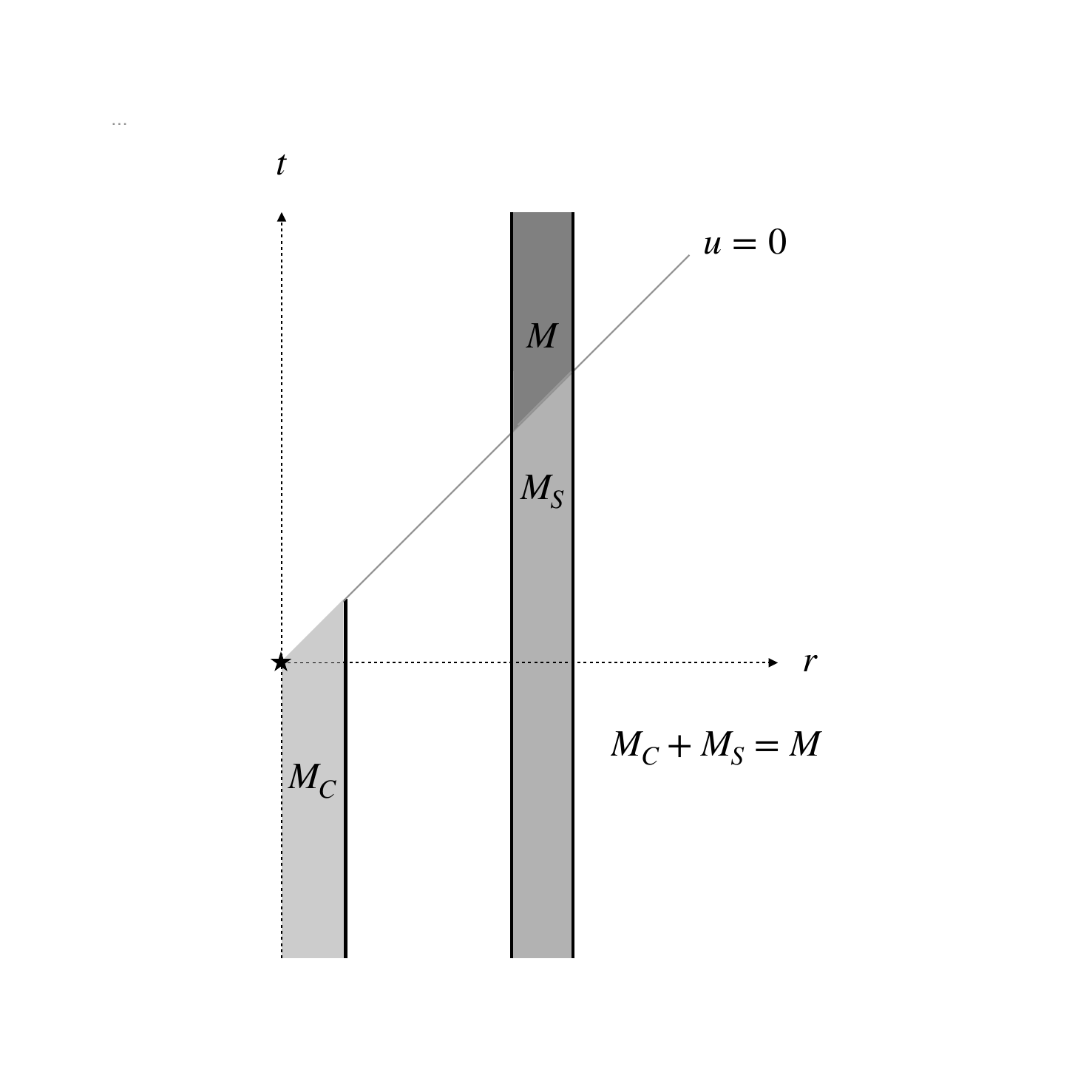} 
\caption{An objective collapse occurring at $r=0$ and $t=0$ localizes all the mass on the shell.}
\end{figure}
%%%%%%%%%%%%%%%%%%%%%%%%%%%%%% 

As for the metric of this SCU, for $u<0$, the metric is given by $ds^2(\rho_C) $. However, for $u \ge 0$ we have $ds^2(0) $. On the $u=0$ surface, the metric is discontinuous, as expected. However, note that the 3-metric induced on it from below and above, $ r^2 d\Omega^2$, is continuous---and, thus, compatible with our gluing condition. 

Before moving on, we point out a potential way of avoiding discontinuities in the metric, which is to introduce an \emph{interpolation} between the metrics before and after the collapse during some short time interval,. For instance, from $u=0$ to $u=\tau_C$, the metric could be given by $\bar{g}=ds^2(\alpha(u))$ with $\alpha(u)$ continuously interpolating between $ \rho_C$ and $0$. It is interesting to note that during the collapse, between the core and the shell, even though $G_{tt}$ remains zero, $G_{tr}$ is different from zero---a fact that could be interpreted as a non-zero radial mass-energy current. In any case, the details of the particular values of the different components of the Einstein tensor depend on the details of the interpolation; and without a well-defined recipe to construct it, this avenue for avoiding discontinuities remains, at best, tentative.

%%%%%%%%%%%%%%%%%%%%%%%%%%%%%%%%%%%%%%%%%%%%%%%%%%%%%%%%%%%%%% 
%%%%%%%%%%%%%%%%%%%%%%%%%%%%%%%%%%%%%%%%%%%%%%%%%%%%%%%%%%%%%%
\section{Tabletop experiments}
\label{Tab}
%%%%%%%%%%%%%%%%%%%%%%%%%%%%%%%%%%%%%%%%%%%%%%%%%%%%%%%%%%%%%% 
%%%%%%%%%%%%%%%%%%%%%%%%%%%%%%%%%%%%%%%%%%%%%%%%%%%%%%%%%%%%%%

The SCU framework outlined in this work, as a semiclassical framework, should lead to empirical predictions manifestly different from approaches assigning a quantum nature to gravity. In line with this, one could look for experimental tests to explore whether spacetime behaves as a classical entity or if it allows for some quantum characteristics. In section \ref{SCG}, we reviewed some of the thought experiments intended to rule out the semiclassical frameworks, but we explained why they fail. More recently, however, there have been proposals that look for a possible quantum behavior of gravity in tabletop experiments. In particular, \cite{MarlettoVedral2017} and \cite{Bose2017}, propose protocols to `witness' the quantum nature of the gravitational field.

These protocols consider two separated quantum systems, $Q_1$ and $Q_2$, such that these systems can only interact if mediated by their gravitational field. It is then alleged that, if the quantum systems are initially in separable states, but they end up entangled, then gravity must have quantum features. If so, a manifestation of entanglement in these experiments would signal a quantum nature of gravity. It is interesting to analyze what would be the predictions of our SCU framework in such experiments.

For concreteness, we consider the proposal in \cite{Bose2017}, where the quantum systems in question are two particles with spin, placed side by side, each in a superposition of two spatial states $\ket{L}_i$ and $\ket{R}_i$. The particles are allowed to interact only by their mutual gravitational interaction. If it is assumed that the gravitation field is quantum, in the sense that the field produced by a particle in a superposition is a superposition of the fields produced by each component, then it can be shown that, during the time interval $\tau$ in which both particles are in their respective superposed states, their gravitational interaction produces a phase shift to the different components of the superposition, depending on their separation. As a result, the gravitational interaction would lead to a state in which the two, initially-separable particles, end up entangled. 

In particular, if the initial state is
\begin{equation}
	\ket{\Phi(0)}=\frac{1}{\sqrt{2}}(\ket{L}_1+\ket{R}_1)\frac{1}{\sqrt{2}}(\ket{L}_2+\ket{R}_2),
\end{equation} 
with $d$ the distance between the centers of the superpositions and $\Delta x$ the distance between the $\ket{L}$ and $\ket{R}$ components, after a time $\tau$, the Newtonian gravitational interaction leads the state to
\begin{equation}\label{BoseEntangledState}
	\ket{\Phi(\tau)}=\frac{e^{i\phi}}{\sqrt{2}}\{\ket{L}_1\frac{1}{\sqrt{2}}(\ket{L}_2+e^{i\Delta\phi_{LR}}\ket{R}_2)+\ket{R}_1\frac{1}{\sqrt{2}}(e^{i\Delta\phi_{RL}}\ket{L}_2+\ket{R}_2)\},
\end{equation} 
with $\Delta\phi_{LR}=\phi_{LR}-\phi$ and $\Delta\phi_{RL}=\phi_{RL}-\phi$, where
\begin{equation}
	\phi\sim\frac{Gm_1m_2 \tau}{\hbar d}, \ \ \ \phi_{LR}\sim\frac{Gm_1m_2 \tau}{\hbar (d+\Delta x)}, \ \ \ \phi_{RL}\sim\frac{Gm_1m_2 \tau}{\hbar (d-\Delta x)}.
\end{equation} 
One can readily see that Eq. (\ref{BoseEntangledState}) represents an entangled state whenever
\begin{equation}
 (\ket{L}_2+e^{i\Delta\phi_{LR}}\ket{R}_2)\neq(e^{i\Delta\phi_{RL}}\ket{L}_2+\ket{R}_2).
\end{equation} 

There are two distinct reasons why, according to our SCUs framework, one would doubt the generation of entanglement in the experimental settings considered. First, according to collapse models, the survival time of a quantum superposition of different localized states depends on the size of the quantum system and those times might be too short compared with the characteristic $\tau$ of experiments that are technologically feasible in the near future. In fact, in \cite{Bose2017} it is acknowledged that, for realistic realizations of the experiment, actual collapse times would inhibit the appearance of gravitationally mediated entanglement.

Second, and more importantly, the semiclassical nature of the SCU framework implies that, for any state of matter---and, in particular, for a state involving a quantum superpositions of relatively well-localized states---there would be a single well-defined spacetime metric and thus, no path for the generation of quantum entanglement. That is, the gravitational field produced by, say, a massive particle in spatial superposition, arises from its corresponding energy-momentum density, $\mathcal{T}_{ab}(x)$, which corresponds to a mass density divided into both components of the superposition. As a result, no gravitationally mediated entanglement is possible.

Still, according to the SCUs framework, these experimental settings would function in a similar fashion to the famous COW experiment \cite{colella1975observation}, where a quantum test particle in a superposition of different localizations is subjected to a classical gravitational field generated by another matter distribution (in that case, the Earth). From the point of view of our framework, the experiments discussed above can be seen as more complicated COW settings, in the sense that each particle is subjected to a ``classical'' gravitational field sourced by the energy-momentum density of the other.
 
Concretely, according to our framework, the initial state above would evolve into
\begin{equation}
	\ket{\Phi(\tau)}_{\text{SCU}} =\frac{1}{\sqrt{2}}(e^{i\alpha}\ket{L}_1+\ket{R}_1)\frac{1}{\sqrt{2}}(\ket{L}_2+e^{i\alpha}\ket{R}_2).
\end{equation} 
with $\alpha=\frac{Gm_1 m_2\tau\ }{\hbar}(\frac{1}{d-\Delta x}-\frac{1}{d+\Delta x})$. That is, no entanglement would be produced, but particular relative phases on each side of the experiment, which could be measured, are predicted by our framework.

%%%%%%%%%%%%%%%%%%%%%%%%%%%%%%%%%%%%%%%%%%%%%%%%%%%%%%%%%%%%%% 
%%%%%%%%%%%%%%%%%%%%%%%%%%%%%%%%%%%%%%%%%%%%%%%%%%%%%%%%%%%%%%
\section{Conclusions}
\label{Con}
%%%%%%%%%%%%%%%%%%%%%%%%%%%%%%%%%%%%%%%%%%%%%%%%%%%%%%%%%%%%%% 
%%%%%%%%%%%%%%%%%%%%%%%%%%%%%%%%%%%%%%%%%%%%%%%%%%%%%%%%%%%%%%

A successful theory of semiclassical gravity---one in which the classical spacetime geometry, described by the Einstein tensor, is coupled to the expectation value of the energy-momentum tensor of quantum matter fields---would represent a significant milestone in the ongoing effort to unify general relativity and quantum mechanics. Even if such a framework does not constitute a full theory of quantum gravity, it could nonetheless serve as an important intermediary step, offering insights into the interface between gravitational and quantum realms.

Despite this potential, the development of a viable semiclassical gravity theory has proven remarkably challenging. To date, no version of such a theory has been both theoretically coherent and consistent with experimental observations. Moreover, the very foundations of semiclassical gravity have been the subject of considerable scrutiny. Over the years, researchers have repeatedly raised concerns about the internal consistency of a framework that attempts to couple fundamentally classical and quantum entities, with several thought experiments and theoretical arguments casting doubt on its validity.

In this work, we address these longstanding issues by proposing a novel framework for semiclassical gravity that is both internally self-consistent and empirically viable. Central to our approach is the incorporation of relativistic objective collapse models within the quantum field theory. In our formulation, the matter fields evolve according to a modified quantum dynamics that includes spontaneous collapse events, providing a clear mechanism for the emergence of definite outcomes. This evolution leads to an expectation value of the energy-momentum tensor that is well-defined at each point in spacetime and suitable for coupling to the classical Einstein tensor.

We have outlined the general principles underlying our framework and provided a detailed example to illustrate its practical implementation. We also explored some of the potential empirical implications of our approach, including possible observational signatures that could distinguish our model from both standard semiclassical theories and full quantum gravity scenarios. This work thus offers a concrete and testable path forward in the ongoing effort to reconcile the quantum and gravitational domains.

%%%%%%%%%%%%%%%%%%%%%%%%%%%%%%%%%%%%%%%%%%%%%%%%%%%%%%%%%%%%%%
\appendix
\section{Spacetime of the example}
\label{EST}
%%%%%%%%%%%%%%%%%%%%%%%%%%%%%%%%%%%%%%%%%%%%%%%%%%%%%%%%%%%%%%
%%%%%%%%%%%%%%%%%%%%%%%%%%%%%%%%%%%%%%%%%%%%%%%%%%%%%%%%%%%%%%

The metric of all static, spherically symmetric spacetimes can be written as
\begin{equation}
ds^2 = -e^{2 \phi(r)} dt^2 + \left(1-\frac{2m(r)}{r} \right)^{-1} dr^2 + r^2 d\Omega^2 ,
\end{equation}
with 
\begin{equation}
m(r) = 4 \pi \int_0^r \rho(r') r'^2 dr'
\end{equation}
and
\begin{equation}
\frac{d\phi}{dr} = \frac{m(r) + 4 \pi r^3 P(r)}{r [r-2 m(r)]} ,
\end{equation}
where $P(r)$ is the radial pressure. Since we are not assuming a perfect fluid, this radial pressure need not be equal to the ``angular pressure'' $F(r)$, given by
\begin{equation}
F(r)= \frac{\rho(r) \left(m(r)+4 \pi r^3 P (r)\right)+P (r) \left(4 \pi r^3 P (r)+2 r -3 m(r) \right)}{2 (r-2 m(r))}+\frac{1}{2} r P'(r)
\end{equation}
Below we consider the different regions in Figure 1.

%%%%%%%%%%%%%%%%%%%%%%%%%%%%%%
\subsection*{Exterior region}
%%%%%%%%%%%%%%%%%%%%%%%%%%%%%%

In the exterior region, the metric is given by
\begin{equation}
ds^2 = -\left(1 - \frac{2M}{r} \right) dt^2 + \left(1-\frac{2M}{r} \right)^{-1} dr^2 + r^2 d\Omega^2,
\end{equation}
i.e., it is the Schwarzschild metric with mass $M$.

%%%%%%%%%%%%%%%%%%%%%%%%%%%%%%
\subsection*{Shell}
%%%%%%%%%%%%%%%%%%%%%%%%%%%%%%

Within the shell,
\begin{eqnarray}
m_S(r) & = & M_C + \frac{4 \pi}{3} \rho_S (r^3 - R_I^3) , \\
\phi_S(r) & = & A_S + \int \frac{m_S(r) + 4 \pi r^3 P(r)}{r [r-2 m_S(r)]} dr
\end{eqnarray}
with $A_S$ such that 
\begin{equation}
e^{2 \phi_S(R_E)} = \left(1 - \frac{2M}{R_E} \right) 
\end{equation}
and, to ensure continuity of $P(r)$ and $F(r)$ at $R_I$ and $R_E$, $P(r)$ satisfying
\begin{eqnarray}
P(R_I) & = & 0, \\
P(R_E) & = & 0, \\
\left. \left( \frac{dP}{dr} \right) \right\rvert_{R_I^+} & = & \frac{M_C \rho_S}{\left(2 M_C - R_I \right) R_I} , \\
\left. \left( \frac{dP}{dr} \right) \right\rvert_{R_E^-} & = & \frac{M \rho_S}{\left(2 M - R_E \right) R_E} .
\end{eqnarray}

%%%%%%%%%%%%%%%%%%%%%%%%%%%%%%
\subsection*{Interior region}
%%%%%%%%%%%%%%%%%%%%%%%%%%%%%%

In the interior region, the metric is given by
\begin{equation}
ds^2 = - e^{2 \phi_S(R_I)} \left(1 - \frac{2M_C}{R_I} \right)^{-1} \left(1 - \frac{2M_C}{r} \right) dt^2 + \left(1-\frac{2M_C}{r} \right)^{-1} dr^2 + r^2 d\Omega^2 .
\end{equation}

%%%%%%%%%%%%%%%%%%%%%%%%%%%%%%
\subsection*{Core}
%%%%%%%%%%%%%%%%%%%%%%%%%%%%%%
Within the core, 
\begin{eqnarray}
m_C(r) & = & \frac{4 \pi}{3} \rho_C r^3 , \\
\phi_C(r) & = & A_C + \int \frac{m_C(r) + 4 \pi r^3 P(r)}{r [r-2 m_C(r)]} dr .
\end{eqnarray}
with $A_C$ such that 
\begin{equation}
e^{2 \phi_C(R_C)} = e^{2 \phi_S(R_I)} \left(1 - \frac{2M_C}{R_I} \right)^{-1} \left(1 - \frac{2M_C}{R_C} \right) 
\end{equation}
and, to ensure continuity of $P(r)$ and $F(r)$ at $R_C$, $P(r)$ satisfying
\begin{eqnarray}
P(R_C) & = & 0, \\
\left. \left( \frac{dP}{dr} \right) \right\rvert_{R_C^-} & = & \frac{M_C \rho_C}{\left(2 M_C - R_C \right) R_C} .
\end{eqnarray}

%%%%%%%%%%%%%%%%%%%%%%%%%%%%%%%%%%%%%%%%%%%%%%%%%%%%%%%%%%%%%% 
%%%%%%%%%%%%%%%%%%%%%%%%%%%%%%%%%%%%%%%%%%%%%%%%%%%%%%%%%%%%%%
\section*{Acknowledgments}
%%%%%%%%%%%%%%%%%%%%%%%%%%%%%%%%%%%%%%%%%%%%%%%%%%%%%%%%%%%%%%
%%%%%%%%%%%%%%%%%%%%%%%%%%%%%%%%%%%%%%%%%%%%%%%%%%%%%%%%%%%%%%

We acknowledge support from PAPIIT grant IG100124. DS acknowledges support for a sabbatical year from UNAM, through a PASPA/DGAPA Fellowship and the hospitality of the Departament de Física Quàntica i Astrofísica, Universitat de Barcelona.

%%%%%%%%%%%%%%%%%%%%%%%%%%%%%%%%%%%%%%%%%%%%%%%%%%%%%%%%%%%%%%
%%%%%%%%%%%%%%%%%%%%%%%%%%%%%%%%%%%%%%%%%%%%%%%%%%%%%%%%%%%%%%
%\bibliographystyle{apalike}
\bibliographystyle{plain}
\bibliography{bibSCU.bib}
%%%%%%%%%%%%%%%%%%%%%%%%%%%%%%%%%%%%%%%%%%%%%%%%%%%%%%%%%%%%%%
%%%%%%%%%%%%%%%%%%%%%%%%%%%%%%%%%%%%%%%%%%%%%%%%%%%%%%%%%%%%%%	 

\end{document}